\documentclass[superbib,superscriptaddress,twocolumn,
nofootinbib,letterpaper,amsmath,amssymb]{revtex4}

\usepackage{graphicx}
\usepackage{subfigure}
\usepackage{amssymb,amsmath,bm}
\usepackage{url}

\usepackage{color}

\usepackage[normalem]{ulem}

\begin{document}

\title{On the security of a new image encryption scheme based on chaotic map lattices}

\author{David Arroyo}
\email{david.arroyo@iec.csic.es}
\affiliation{Instituto de
F\'{\i}sica Aplicada, Consejo Superior de Investigaciones
Cient\'{\i}ficas, Serrano 144---28006 Madrid, Spain}
\author{Rhouma Rhouma}
\affiliation{Syscom Laboratory, Ecole Nationale d'Ing\'{e}nieurs de
Tunis, 37, Le Belv\'{e}d\`{e}re 1002 Tunis, Tunisia}
\author{Gonzalo Alvarez}
\affiliation{Instituto de F\'{\i}sica Aplicada, Consejo Superior de
Investigaciones Cient\'{\i}ficas, Serrano 144---28006 Madrid, Spain}
\author{Shujun Li}
\affiliation{FernUniversit\"{a}t in Hagen, Lehrgebiet
Informationstechnik, Universit\"{a}tsstra{\ss}e 27, 58084 Hagen,
Germany}
\author{Veronica Fernandez}
\affiliation{Instituto de F\'{\i}sica Aplicada, Consejo Superior de
Investigaciones Cient\'{\i}ficas, Serrano 144---28006 Madrid, Spain}

\begin{abstract}
This paper reports a detailed cryptanalysis of a recently proposed
encryption scheme based on the logistic map. Some problems are
emphasized concerning the key space definition and the
implementation of the cryptosystem using floating-point operations.
It is also shown how it is possible to reduce considerably the key
space through a ciphertext-only attack. Moreover, a timing attack
allows the estimation of part of the key due to the existent
relationship between this part of the key and the
encryption/decryption time. As a result, the main features of the
cryptosystem do not satisfy the demands of secure communications.
Some hints are offered to improve the cryptosystem under study
according to those requirements.
\end{abstract}


\maketitle

\bfseries

Recently a new cryptosystem was proposed by using a chaotic map
lattice (CML). In this paper, we analyze the security of this
cryptosystem and point out some of its security defects. A number of
measures have been suggested to enhance the security of the
cryptosystem following some established guidelines on how to design
good cryptosystems with chaos.

\mdseries

\section{Introduction}

Image encryption is somehow different from text encryption due to
some inherent features of images, such as bulk data capacity and
high correlation among pixels. Therefore, digital chaotic ciphers
like those in \cite{bap,li1,alve} and traditional cryptographic
techniques such as DES, IDEA and RSA are no longer suitable for
practical image encryption, especially for real-time communication
scenarios. So far, many chaos-based image cryptosystems have been
proposed \cite{chen,guan,pareek,kwok,rhchaoss}. The major core of
these encryption systems consists of one or several chaotic maps
serving the purpose of either just encrypting the image or shuffling
the image and subsequently encrypting the resulting shuffled image.
In \cite{pisarchik06} a new image encryption algorithm based on
chaotic map lattices has been proposed. The aim of this paper is to
assess the security of such cryptosystem.

The rest of the paper is organized as follows.
Section~\ref{section:description} describes the cryptosystem
introduced in \cite{pisarchik06}. After that,
Section~\ref{section:DesignProblems} points out some design problems
inherent to that cryptosystem, and Section~\ref{section:attacks}
gives some attacks on the cryptosystem under study. Finally, some
security enhancements are presented in Section~\ref{section:enhance}
followed by the last section, which presents the conclusions.

\section{Description of the encryption scheme}
\label{section:description}

The encryption scheme described in \cite{pisarchik06} is based on
the logistic map given by
\begin{equation}
x_{i+1}=a \cdot x_i \cdot (1-x_i).\label{equation:logistic}
\end{equation}
For a certain value of $a$, the chaotic phase space is
$[x_{\min},x_{\max}]$.

Given an $M\times N$ color image with R, G, B color components, an
initialization process is performed to convert the integer values of
each pixel to real numbers that can work with the above chaotic
logistic map. First, the 2-D image is scanned in the raster order
(i.e., from left to right and from top to bottom) to form three 1-D
integer sequences $\{P_c^i\}_{i=1}^m$ ($c=\text{R, G and B}$), where
$P_c^i\in\{0,\cdots,255\}$ denotes the color component $c$ of the
$i$-th pixel. Then, the integer sequences are converted to three
real-number sequences each of which corresponds to a different color
component: $\{x_c^i(0)\}_{i=1}^m$, where
\begin{equation}
x_c^i(0)=x_{\min}+(x_{\max}-x_{\min})\cdot
P_c^i/255.\label{eq:integer2real}
\end{equation}
After the above initialization process, the following encryption
procedure is carried out separately for each color component to
obtain the ciphertext:

\begin{enumerate}
\item
Set $r=1$.

\item
Set the initial condition of the logistic map as follows:
\[
x_0=\begin{cases}%
x_c^m(r-1), & \mbox{if }i=1,\\
x_c^{i-1}(r), & \mbox{if }2\leq i\leq m.
\end{cases}
\]
\label{algorithm:encryption1}

\item
Iterate the chaotic logistic map from $x_0$ for $n$ times to obtain
$x_n$. \label{algorithm:encryption2}

\item
Set $x_c^i(r)=x_n+x_c^i(r-1)$. If $x_c^i(r)>x_{\max}$, then subtract
$(x_{\max}-x_{\min})$ from $x_c^i(r)$ to ensure
$x_c^i(r)\in[x_{\min},x_{\max}]$.

\item
Set $r=r+1$. If $r<j$, go to Step~\ref{algorithm:encryption1};
otherwise the encryption procedure stops for the current color
component.
\end{enumerate}

After performing the above encryption procedure for all three color
components, the three sequences $\{x_R^i(j)\}_{i=1}^m$,
$\{x_G^i(j)\}_{i=1}^m$ and $\{x_B^i(j)\}_{i=1}^m$ make up the
ciphertext.

As claimed in \cite{pisarchik06}, the secret key is composed of the
following four sub-keys:
\begin{enumerate}
\item
The control parameter of the logistic map, i.e., $a$.

\item
The image height and the image width, i.e., $M$ and $N$
respectively.

\item
The number of chaotic iterations in
Step~\ref{algorithm:encryption2}, i.e., $n$.

\item
The number of cycles, i.e., $j$.
\end{enumerate}

The decryption procedure is similar to the above description, but in
an reverse order, and the following inverse map
\begin{equation}
P_c^i=\text{round}[(x_c^i(0)-x_{\min})\cdot
255/(x_{\max}-x_{\min})]\label{eq:real2integer}
\end{equation}
is used in the last step to recover the plain-image by converting
real numbers back to integer pixel values. For more details about
the encryption/decryption procedures, the reader is referred to
\cite{pisarchik06}.

\section{Design problems}
\label{section:DesignProblems}

\subsection{Key definition problems}

Following Kerckhoffs' principle \cite{menezes:book97}, the security
of a cryptosystem should depend only on its key. For the
cryptosystem defined in \cite{pisarchik06}, the size of the image to
be encrypted determines one of its four secret sub-keys. In a
known-plaintext attack we have access to both the plain image and
its encrypted version, which means that we know the size of the
image. Moreover, in a ciphertext-only attack the value $m=M\times N$
is known and it is possible to get $M$ if $N$ is known and vice
versa. Therefore, it is not a good idea to include the size of the
image as part of the key, since it does not increase the difficulty
to break the cryptosystem.

\begin{figure}
\includegraphics{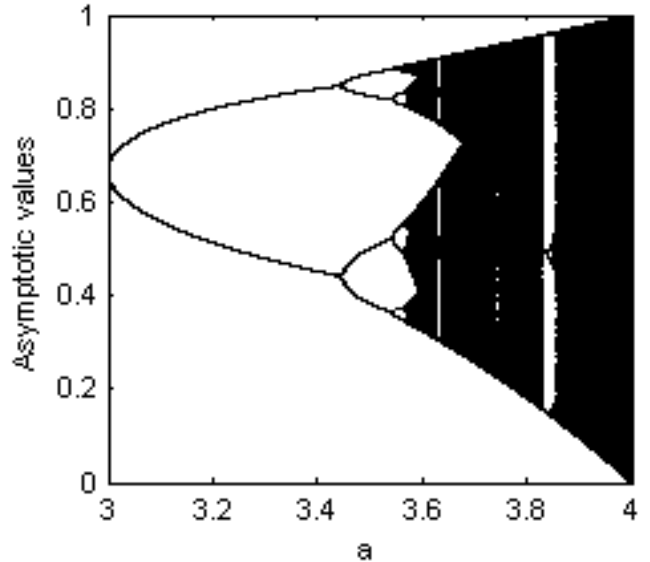}
\caption{Bifurcation diagram of the logistic map showing periodic
windows.} \label{figure:logistic}
\end{figure}

In addition, the control parameter $a$ of the logistic map is also
part of the key. In \cite{pisarchik06} $a$  is chosen in $(3.57,4)$
for the sake of the map defined in Eq. (\ref{equation:logistic})
being always chaotic. However, the bifurcation diagram of the
logistic map (Fig. \ref{figure:logistic}) shows the existence of
periodic windows in that interval. It means that a user could choose
$a$ such that the logistic map would be working in a non-chaotic
area, which is not a good security criterium when considering
chaotic cryptosystems \cite[Rule 5]{Alvarez06a}. Hence, it is
advisable to give a more detailed definition of the possible values
of $a$, so that the user can only choose those values of the control
parameter $a$ preventing the logistic map from showing a periodic
behaviour.

Finally, the other parts of the key are the number of iterations of
the logistic map per pixel ($n$) and the number of encryption cycles
($j$). As secret sub-keys, both values should possess a high level
of entropy to avoid being guessed by a possible attacker. However,
it is not advisable to select large values for $j$ and $n$, since it
will definitely lead to a very slow encryption speed. On the other
hand, using small values of $j$ and $n$ reduces the level of
security, since those small values do not provide good confusion and
diffusion properties. Both restrictions imply a reduction of the
associated sub-key space and thus they make the brute-force attack
more likely to be successful. As a conclusion, it is convenient to
use $j$ and $n$ as design parameters and not as part of the secret
key. This approach has been traditionally followed with respect to
the number of encryption rounds in classical schemes such as DES or
AES.

\subsection{Underlying decryption error}

As it happens during the encryption procedure, all the intermediate
values $x_c^i(r)$ obtained through the decryption stage must be
inside the phase space. This means that $x_{\min}$ should appear in
Eq.~(10) and (11) in \cite{pisarchik06} instead of $0$. Having in
mind this consideration, the performance of the decryption process
will be analyzed in the following.

The cryptosystem described in \cite{pisarchik06} generates a
ciphertext consisting of a number of real values. All the operations
to encrypt an image are performed using floating-point arithmetic.
From Section~\ref{section:description} we know that
$x_c^i(r)=x_n+x_c^i(r-1)$, where $x_n$ is the resulting value of
iterating the logistic map $n$ times from $x_0$. Hence, if we want
to recover $x_c^i(r-1)$ (the original value of the $i$-th element in
the last round), we have to iterate $n$ times the logistic map from
$x_0$ to get $x_n$ and, after that, to substract this value from
$x_c^i(r)$. However, the resulting value of this previous operation
might not match the actual value of $x_c^i(r-1)$, due to the
wobbling precision problem that exists when dealing with
floating-point operations \cite[p. 39]{floatingPoint:higham}. This
wobbling precision problem also causes the resulting guessed value
of $x_c^i(r-1)$ to depend on the cryptosystem implementation.
Therefore, if an image is encrypted on one platform and decrypted on
another, and the implementations of floating-point arithmetics on
both platforms are not compatible with each other, then the
decrypted image might not match the original one. In
\cite{pisarchik06} the cryptosystem was implemented using Microsoft
Visual C\# .NET 2005 and no comment was given about the wobbling
precision problem in the decryption process. However, we have
experimentally verified that this problem indeed exists when the
cryptosystem is implemented using MATLAB. A very useful measure of
the performance of the decryption procedure is the Mean Square Error
or MSE. For $P$ and $P^\prime$ being a plain image and the decrypted
image respectively, the MSE for the color component $c$ is defined
as
\begin{equation}
    MSE_c = \sum(P_c^i-{P^\prime}_c^i)^2/m,
\end{equation}
where $m=M\times N$ is the number of pixels of the images
considered. Consequently, for a well designed encryption/decryption
scheme the MSE should be 0 for each color component. Unfortunately,
for the cryptosystem under study, the values of MSE for all three
color components are generally not equal to 0 due to the wobbling
precision problem associated to the floating-point arithmetic.

\begin{figure}[!htbp]
\includegraphics{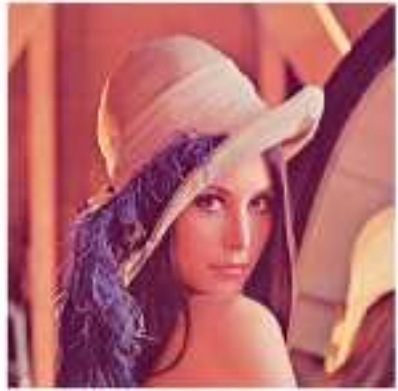}
\caption{The plain-image ``Lena''.} \label{figure:Lena}
\end{figure}

\begin{figure*}[!htbp]
\subfigure[]{\includegraphics{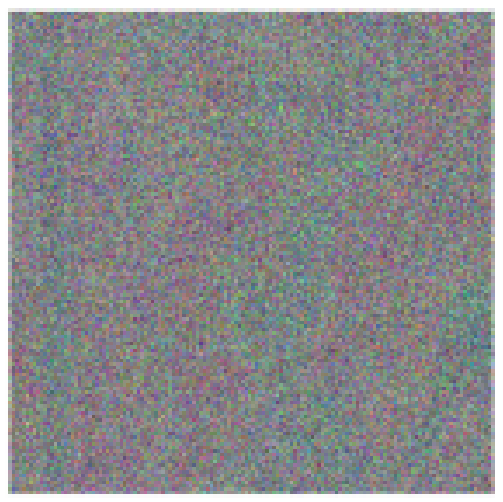}}
\subfigure[]{\includegraphics{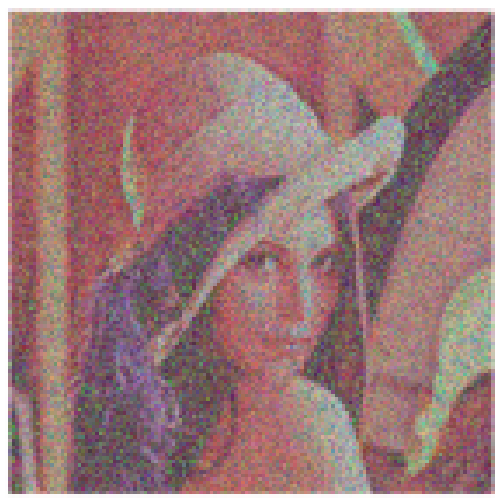}}
\subfigure[]{\includegraphics{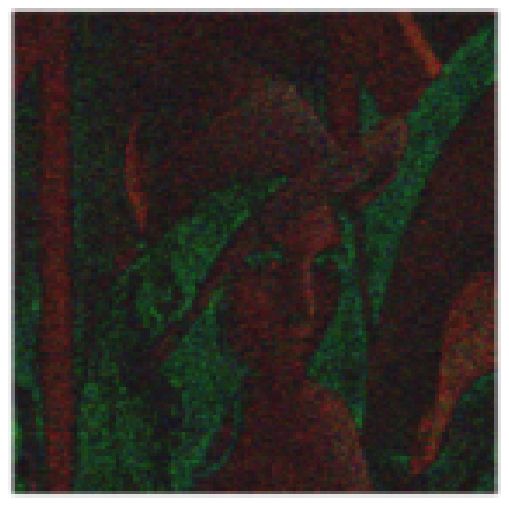}} \caption{Simulations with
MATLAB (a) Ciphertext of the plain-image ``Lena'' (visualized as a
pseudo-image by using Eq.~\eqref{eq:real2integer}) (b) Recovered
image of ``Lena'' using the same key (c) The error image between the
original and the recovered ``Lena''.} \label{fig_error}
\end{figure*}

In order to evaluate the underlying decryption error of the
cryptosystem defined in \cite{pisarchik06}, a $512\times 512$
plain-image ``Lena'', as shown in Fig.~\ref{figure:Lena}, was
encrypted and decrypted using the same key $(n,j,a)=(30,1,3.9)$. The
results showed that the three MSEs obtained for the red, green and
blue components of the decrypted image with respect to the original
one were 6.49, 0.018, 0.057, respectively. For another key
$(n,j,a)=(30,3,3.9)$, the obtained MSEs were 206.96, 123.45, 58.65,
respectively. Figure~\ref{fig_error} shows the decrypted image and
the error image when the cryptosystem was implemented in MATLAB
using a third key $(n,j,a)=(5,2,3.9)$.

\section{Attacks}
\label{section:attacks}

\subsection{Control parameter estimation}
\label{subsection:paramEstimation}

The maximum value of $x_{i+1}$ in Eq.~\eqref{equation:logistic} is
reached when $x_i=0.5$, which informs that the maximum value of a
sequence generated from the iteration of the logistic map is $a/4$,
i.e., $x_{\max}=\max \left(\left\{x_i\right\}\right)\leq a/4$. The
ciphertext of the cryptosystem proposed in \cite{pisarchik06} is
composed of $3m$ real values, each of which is in the range
$[x_{\min},x_{\max}]$. This means that it is possible to approximate
$x_{\max}=a/4$ as the maximum value of all the real values in the
ciphertext, i.e.,
\begin{equation}
\hat x_{\max} = \max_{1\leq i\leq m \atop c=\text{R, G, B}}x_c^i(j).
\label{equation:xmax}
\end{equation}
Then, from $x_{\max}=a/4$, one can estimate the secret value of the
control parameter $a$ as
\begin{equation}
\hat a = 4 \cdot \hat x_{\max}. \label{equation:parameter}
\end{equation}

\begin{figure}
\includegraphics{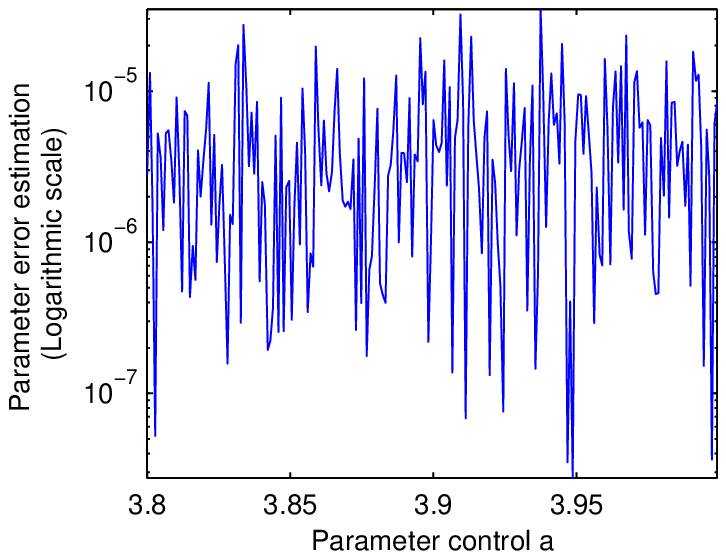}
\caption{Parameter estimation errors corresponding to the image
``Lena'', when $n=20$ and $j=1$.}
\label{figure:parameterEstimationLena}
\end{figure}

\begin{figure}
\includegraphics{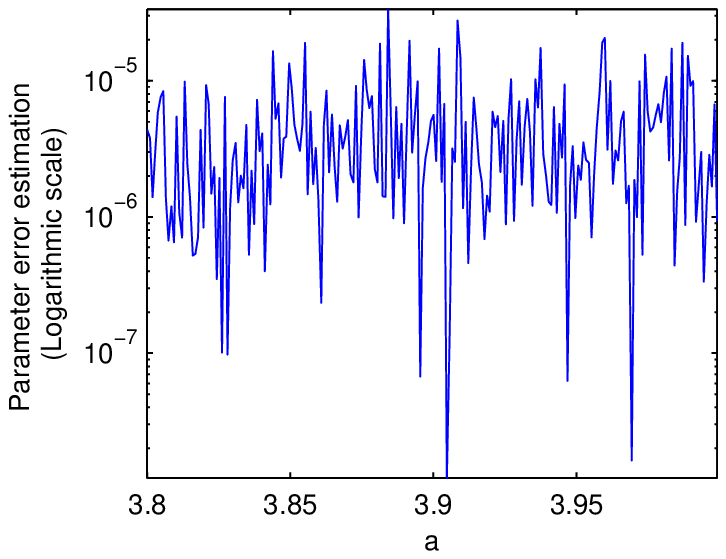}
\caption{Parameter estimation errors corresponding to the image
``Lena'', when $n=20$ and $j=3$.}
\label{figure:parameterEstimationLena_2}
\end{figure}

Consequently, if we have a ciphertext, we can estimate the value of
the sub-key $a$. In other words, a ciphertext-only attack allows us
to estimate the sub-key $a$. In this sense, the image ``Lena''
(Fig.~\ref{figure:Lena}) was encrypted for $n=20$, $j=1$ and
different values of $a\in[3.8,4]$. These values of $a$ were then
estimated from the ciphertexts by applying
Eqs.~\eqref{equation:xmax} and \eqref{equation:parameter}. The
estimation errors are shown in
Fig.~\ref{figure:parameterEstimationLena}. The average estimation
error was $5.236228\times 10^{-6}$, whereas the maximum and minimum
errors were $3.481322\times 10^{-5}$ and $2.758853\times 10^{-8}$,
respectively. By increasing the value of $j$ from 1 to 3 and keeping
the other sub-keys unchanged, the parameter estimation errors are
shown in Fig.~\ref{figure:parameterEstimationLena_2}, being the mean
estimation error $4.721420\times 10^{-6}$, the minimum error
$1.212016\times 10^{-8}$ and the maximum error $3.355227\times
10^{-5}$.

\begin{figure*}[!htbp]
\subfigure[~Red
component]{\includegraphics[scale=0.8]{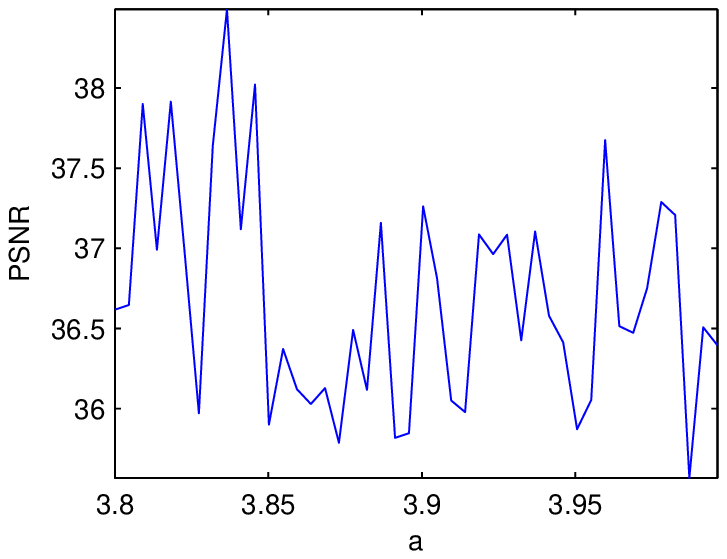}}
\subfigure[~Green
component]{\includegraphics[scale=0.8]{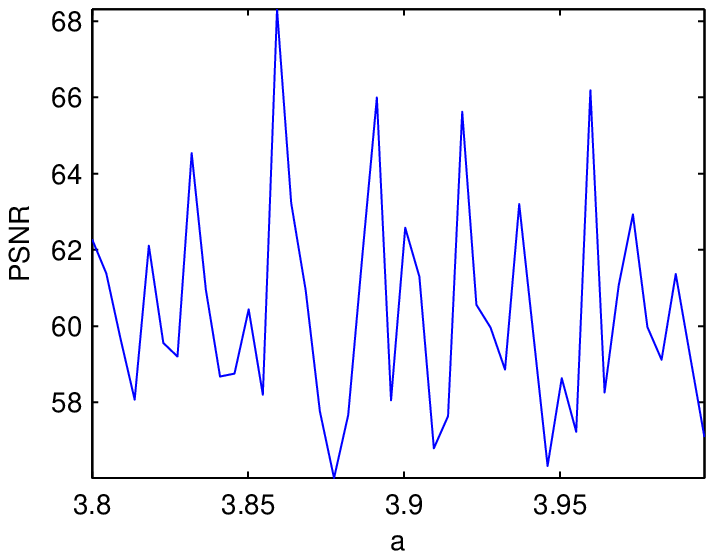}}
\subfigure[~Blue
component]{\includegraphics[scale=0.8]{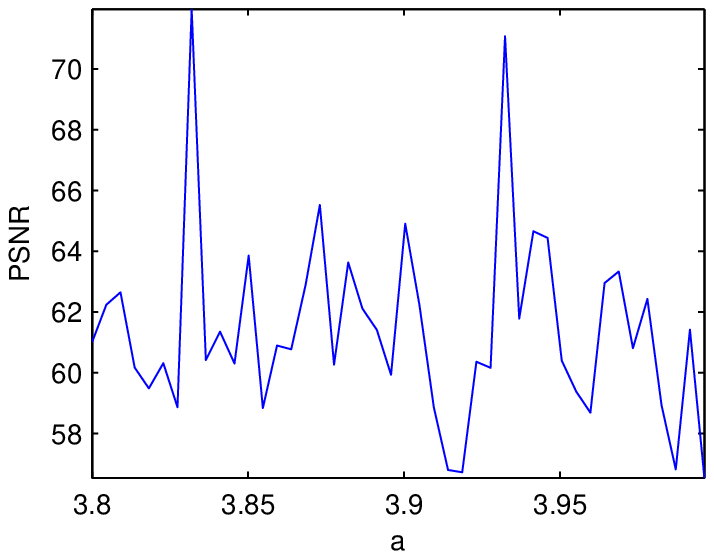}}
\caption{PSNRs of the decrypted image ``Lena'' with respect to
different values of the control parameter $a$.} \label{figure:psnr1}
\end{figure*}

\begin{figure*}[!htbp]
\subfigure[~Red
component]{\includegraphics[scale=0.6]{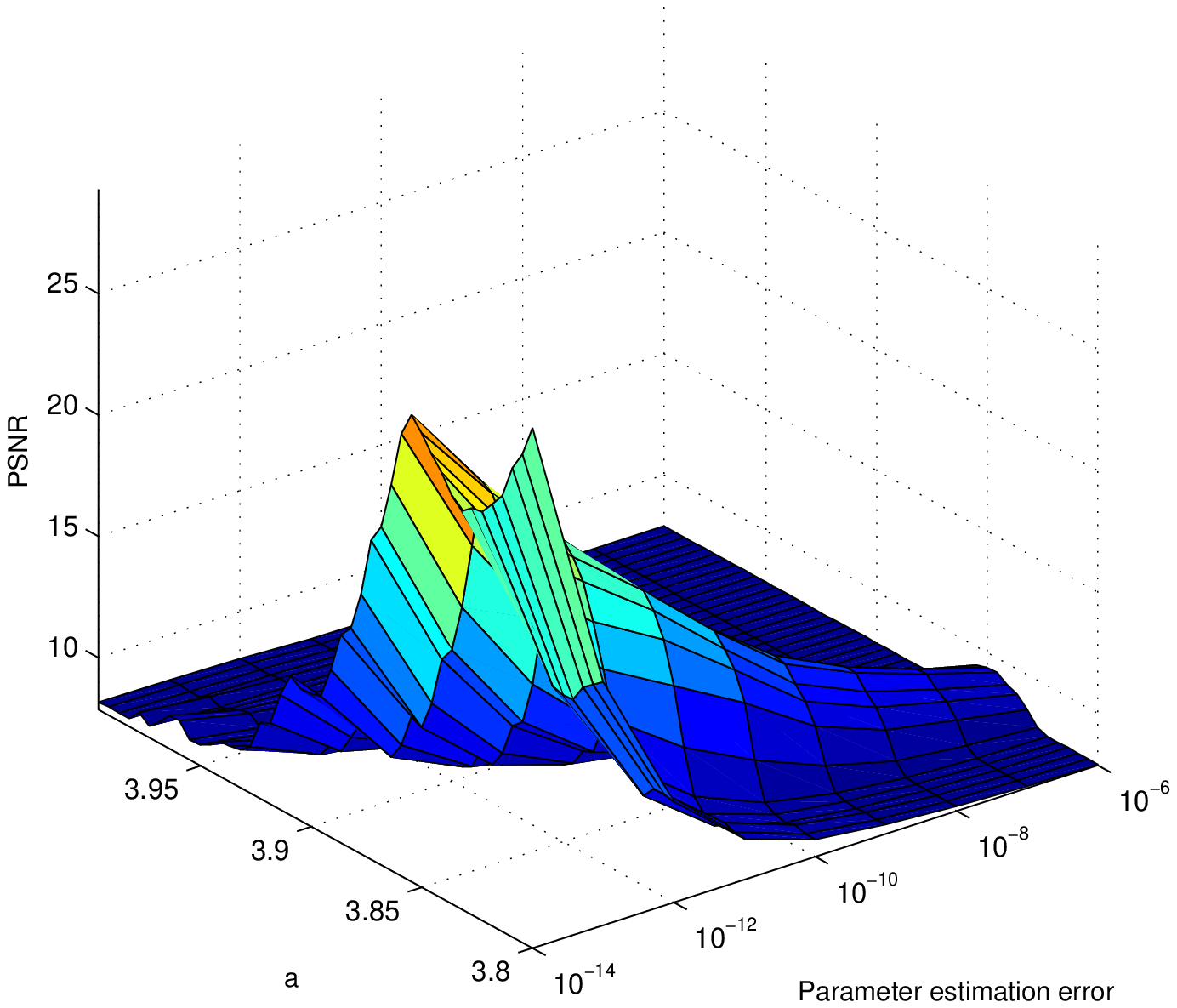}}
\subfigure[~Green
component]{\includegraphics[scale=0.6]{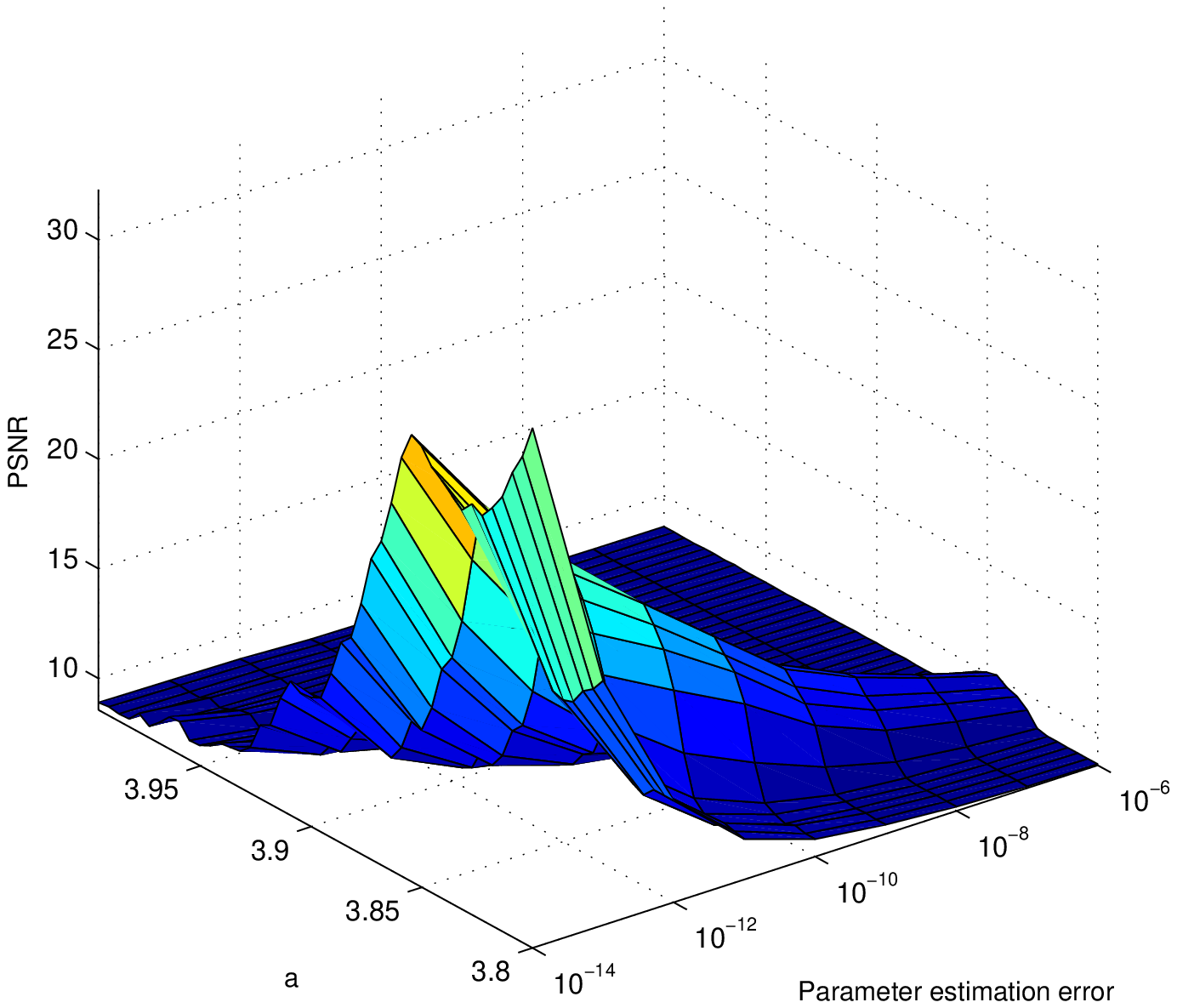}}
\subfigure[~Blue
component]{\includegraphics[scale=0.6]{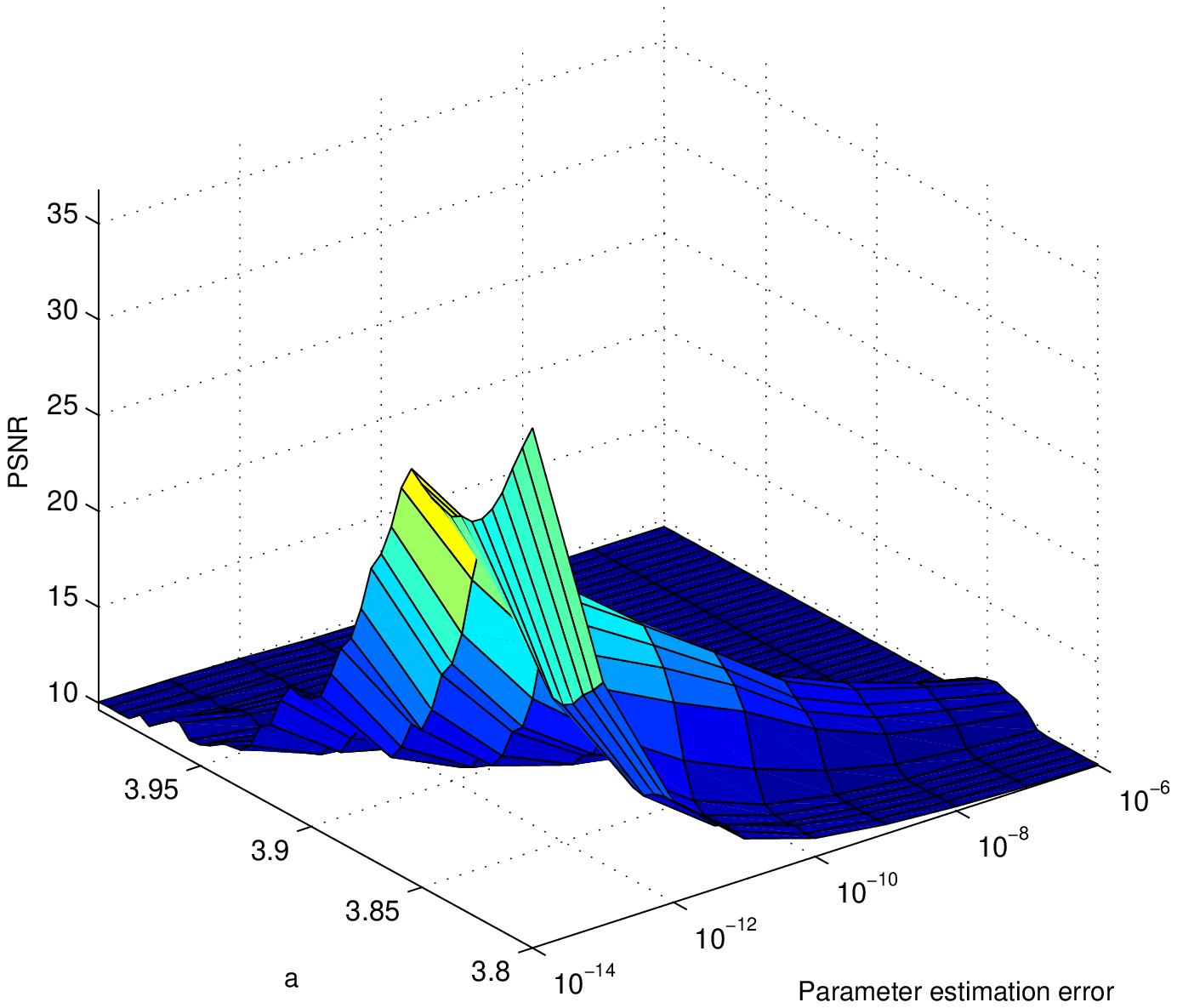}}
\caption{PSNRs of the decrypted image ``Lena'' with respect to
different values of the control parameter $a$ and different
parameter estimation errors.} \label{figure:psnr2}
\end{figure*}

Finally, in Figs.~\ref{figure:psnr1} and \ref{figure:psnr2} the
sensitivity of the cryptosystem with respect to the control
parameter $a$ is shown. This sensitivity is measured using the Peak
Signal to Noise Ratio (PSNR), which is defined for the color
component $c$ as
\begin{equation}
PSNR_c=10\cdot\log_{10}\left(\frac{255^2}{MSE_c}\right).
\end{equation}

Figure~\ref{figure:psnr1} displays the PSNRs of the different color
components of the decrypted image ``Lena'' with respect to the
original image ``Lena'' for $a\in[3.8,4]$ when the same key is used
for encryption and decryption. The values of the other sub-keys are
$n=20$, $j=3$. On the other hand, Figure~\ref{figure:psnr2} shows
the PSNRs when the control parameter used in decryption shows some
deviation from that employed in the encryption process. One can see
that for a deviation of the control parameter of less than
$10^{-10}$ and for a certain range of values of the control
parameter, it is possible to recover the original image ``Lena''
with a similar PSNR to that obtained using the correct control
parameter. For instance, for $a=3.845621$ the PNSRs for the red,
green and blue components of the recovered ``Lena'' are $35.899819$,
$60.437331$ and $63.853450$, respectively. For the same value of $a$
and a parameter estimation error equal to $10^{-12}$, the PSNR of
the recovered ``Lena'' with respect to the original one is
$17.480625$ for the red component, $18.622578$ for the green and
$20.019512$ for the blue component.

\subsection{Timing attack}

One important feature of a secure encryption scheme is that the
encryption speed should not depend on the key value. Indeed, if the
time consumed on encryption/decryption is correlated with the value
of the key (or a sub-key), then it is possible to approximate that
(sub-)key. This kind of attack is called timing attack
\cite{timingAttack:Kocher96, timingAttack:Brumley03}. As it has been
shown in Section~\ref{section:description}, in every encryption
round, Step~\ref{algorithm:encryption2} is carried out through the
$n$ iterations of Eq.~\eqref{equation:logistic}, where $n$ is a
sub-key. This means that, for a certain number of encryption rounds
(i.e., a certain value of $j$) and a certain value of the control
parameter $a$, the encryption speed decreases as $n$ does.
Similarly, because the encryption/decryption procedure is composed
of $j$ repeated cycles, the encryption speed will also become slower
if the value of $j$ increases. To be more precise, for a given
plain-image, we can expect the existence of the following bi-linear
relationship between the encryption/decryption time (EDT) and the
values of $n$ and $j$:
\begin{equation}
   EDT(n,j)\approx (c\times n+d_0)\times j+d_1,
   \label{equation:EDT_approx}
\end{equation}
where $c$ corresponds to the common operations consumed on each
chaotic iteration, $d_0$ to the operations performed in each cycle
excluding those about chaotic iterations, and $d_1$ to those
operations performed on the initialization process and the
postprocessing after all the $j$ cycles are completed. In addition,
because $a$ is just the control parameter of the chaotic map, it is
expected that $EDT$ will be independent of its value.

\begin{figure*}[!htbp]
\includegraphics{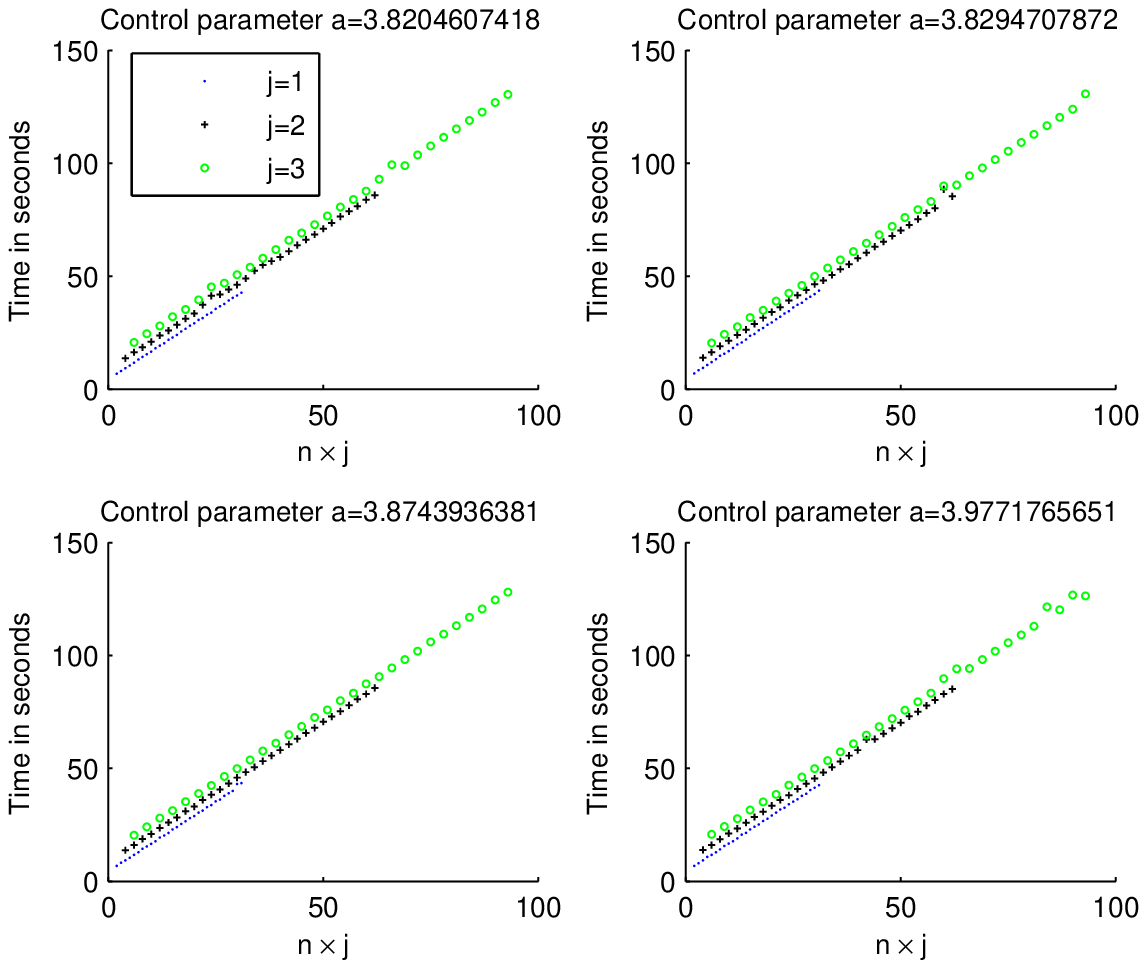}
\caption{Encryption time for images of size $256\times 256$ and
different values of the number of iterations $n$ and the number of
encryption rounds.} \label{figure:encryptionTime}
\end{figure*}

With the aim of verifying this hypothesis, some experiments have
been made under the following scenario. An image with random pixel
values of size $256 \times 256$ was encrypted for different values
of $a$, $n$ and $j$. The encryption time corresponding to each key
is shown in Fig.~\ref{figure:encryptionTime}, from which one can see
that Eq.~\eqref{equation:EDT_approx} is verified.

The above experimental results ensure the feasibility of a timing
attack to a sub-key of the cryptosystem under study: by observing
the encryption time, it is possible to estimate the values of $n$ if
$j$ is known and vice versa. Without loss of generality, assuming an
attacker Eve knows the value of $n$, but not that of $j$, let us
demonstrate how the timing attack can be performed in practice. In
this case, the relationship between EDT and the value of $j$ can be
simplified as $EDT(n,j)=c_n\times j+d_n$, where $c_n=c\times n$ and
$d_n=d_0\times j+d_1$. Then, if Eve gets a temporary access to the
encryption (or decryption) machine, she can carry out a real timing
attack in the following steps:
\begin{enumerate}
\item
She observes the whole process of encryption (or decryption) to get
the encryption (or decryption) time $t_j$ and also the size of the
ciphertext (i.e., the size of the plaintext).

\item
By choosing two keys with different values of $j$, she
encrypts\footnote{Please note that this can be done on her own
computer, as long as she has the encryption/decryption software
installed.} a plaintext (or decrypts a ciphertext) of the same size
and gets $t_1$ and $t_2$.

\item
She derives the values of $c_n$ and $d_n$ by substituting $t_1$ and
$t_2$ into $EDT(n,j)=c_n\times j+d_n$.

\item
She estimates the value of $j$ to be
$\hat{j}=\text{round}((t_j-d_n)/c_n)$.

\item
She verifies the estimated value $\hat{j}$ by using it to decrypt
the observed ciphertext. If the recovered plaintext is something
meaningful, the attack stops; otherwise, she turns to search the
correct value of $j$ in a small neighborhood of $\hat{j}$ until a
meaningful plaintext is obtained.
\end{enumerate}

The above timing attack actually reveals that partial knowledge
about the key constitutes useful information to determine the rest
of the key. However, such a problem should not exist for a
well-designed cryptosystem \cite[Rule~7]{Alvarez06a}. Hence, we
reach the conclusion that the cryptosystem proposed in
\cite{pisarchik06} was not well designed.

Finally, it deserves being mentioned that the linear relationship
between the encryption/decryption time and the value of $j$ has been
implicitly shown in \cite[Table~I]{pisarchik06}. There, for an image
of size $300 \times 200$ and $j$ equal to $1$, $2$ and $3$, the
encryption times were observed to be $13.6$, $26.7$ and $39.1$
seconds, respectively. This clearly showed a linear relationship
between the encryption time and the value of $j$. Unfortunately, the
authors of \cite{pisarchik06} did not realize that this is a
security defect that could be used to develop the timing attack
reported in this paper.

\section{Enhancements}
\label{section:enhance}

To overcome the problems of the original cryptosystem, we propose to
enhance it by applying the following rules:
\begin{itemize}
\item
Use a piecewise linear chaotic map (PWLCM) \cite{Li:DPWLCM:IJBC2005}
instead of the logistic map for the size of the chaotic phase space
being independent with respect to the control parameter value.
Indeed, the chaotic phase space of the PWLCM is (0,1) for all the
values of the control parameter. The PWLCM also has a uniform
invariant probability distribution function, which makes impossible
to estimate the control parameter through the maximum value of the
ciphertext, as we can do for the cryptosystem under study.

\item
The wobbling precision problem should be circumvented by forcing
fixed-point computations. A possible solution is to transform the
values of the phase space of the chaotic map into integer values, so
the encryption and decryption operations are carried out using
integer numbers instead of real numbers.

\item
Without loss of security, the enhanced cryptosystem should be easy
to implement with acceptable cost and speed
\cite[Rule~3]{Alvarez06a}. It is expected that the enhanced
cryptosystem can encrypt at least a pixel per iteration to reach
high encryption/decryption speed.

\item
The key of the enhanced cryptosystem should be precisely defined
\cite[Rule~4]{Alvarez06a}, and the key space from which valid keys
are chosen should be precisely specified and avoid non-chaotic
regions \cite[Rule~5]{Alvarez06a}. This can be assured by choosing
the control parameter(s) of a PWLCM as the secret key, because for
every valid control parameter, the behavior of the PWLCM is chaotic.

\item
Having in mind today's computer speed, the key space size should be
$\kappa > 2^{100}=10^{30}$ in order to elude brute-force attacks
\cite[Rule~15]{Alvarez06a}. In the encryption scheme defined in
\cite{pisarchik06} every color component is encrypted independently
from the other color components. Nevertheless, the secret key
employed in the encryption process of each color component is the
same. It is convenient to use a different value of the key for each
color component and make the encryption of the three color
components dependent on each other, since this implies a
considerable increase of the key space. It has been tested that the
sensitivity of the PWLCM with respect to the control parameter is
around $10^{-10}$. Therefore, when the control parameter is used as
the key of the cryptosystem, the size of the key space will be
$\kappa =10^{10}$. Nonetheless, if we use a different value of $p$
for every color component, and the encryption of each color
component depends on the others, the size of the key space will be
$\kappa = 10^{30}$, which satisfies the security requirement related
to the resistance against brute-force attacks.
\end{itemize}

\section{Conclusions}
\label{section:conclusion}

In this paper, some problems of a new image encryption scheme based
on chaotic map lattices are reported and two attacks on this
cryptosystem have been presented. To overcome these problems and
weaknesses, we have introduced some countermeasures to enhance the
cryptosystem by following the cryptographical rules listed in
\cite{Alvarez06a}.

\section{Acknowledgments}
The work described in this paper was partially supported by
\textit{Ministerio de Educaci\'on y Ciencia of Spain}, research
grant SEG2004-02418 and \textit{Ministerio de Ciencia y
Tecnolog\`{i}a} of Spain, research grant TSI2007-62657. Shujun Li
was supported by a research fellowship from the \textit{Alexander
von Humboldt Foundation, Germany}.

\newpage
\bibliographystyle{chaos}

\end{document}